\documentclass{roajarticle}
\usepackage{graphicx}
\usepackage{mynatbib}
\usepackage{upgreek}
\usepackage{upgreek2}
\usepackage{amsmath}
\usepackage[section]{placeins}
\newcommand{\volume}{\textbf{1}}
\newcommand{\numarul}{1}
\newcommand{\volyear}{2019}
\graphicspath{ {./figs/} }

\title{Solar eclipse observations with small radio telescope in Hong Kong in 21cm radio frequency band}
\newcommand{\shorttitle}{Solar eclipse observations}

\author[1]{C. S. Leung}

\author[2]{Thomas K. T. Fok}
\author[2]{ Kenneith H. K. Hui}
\author[3]{K. W. Ng}
\author[3]{C. M. Lee}
\author[3]{S. H. Chan}
\affil[1]{Department of Applied Mathematics, Hong Kong Polytechnic University, Hong Kong SAR, P. R. China,
Email: chun-sing-hkpu.leung@polyu.edu.hk }
\affil[2] {	
Ho Koon Nature Education cum Astronomical Centre , Sik Sik Yuen, Hong Kong SAR, P. R. China}
\affil[3] {Hong Kong Astronomical Society, Hong Kong SAR, P. R. China}
\keywords{Solar observations}

\begin{document}
\maketitle

\begin{abstract}
Small radio telescope in 21cm was used for studying the partial solar eclipse, with magnitude 0.89, in Hong Kong on 21st June, 2020. The radio telescope SPIDER 300A was designed and constructed by the Radio2Space Company, Italy. Radio flux density time curves (light curve) and a two-dimension mapping of the eclipse is presented in this paper. Standard radio data reduction methods were used to obtain the intensity time curve. We also adopted the semi-pipeline method for the reduction of data to obtain the same results as with the built-in software of the radio telescope SPIDER 300A. The total solar radio flux of the eclipse was found to reduce by maximum 55 $\pm$ 5 percent, while the maximum eclipsed area of the same eclipse is 86.08\%.
Other radio observations of solar eclipses in Hong Kong are also discussed in this paper, including SPIDER 300A observation of partial solar eclipse on 26th December 2019 (APPENDIX A); and small radio telescope (SRT), developed by the Haystack Observatory, MIT, USA, observation of 2020 eclipse (APPENDIX B).
\end{abstract}

\section{Introduction}
The first radio solar physics paper was published at 1944 \citep{Reber1944}. After that, solar radio physics was born. Following this development, several interesting books with details for introducing in solar radio physics keep momentum in the field \citep{Kundu1965, Kruger1979, Kundu1980, McLean1985}.

  For solar radio astronomy studies, total or partial eclipses provide great opportunities for solar radio observations. Solar radio observation primarily are time-dependent. Furthermore, several radio solar eclipse observations were carried out in China in 1958, 1968, 1980 and 1987. And the findings were briefly introduced \citep{Liu1998}.

  On November 3 1994, high spectral and time resolution radio observations of solar eclipse were carried out at Chapeco, Brazil  for the first time \citep{Sawant1997}, asymmetrical limb brightening at 1.5 GHz was observed and recorded. Besides, thanks to the advance in technology, sensitivity for the backend system had improved a lot so that two-dimensional mapping for the radio sun can easily be achieved. Recently, one of the interesting results was published by making use a UK made telescope at University of Baghdad \citep{Jallod2019}. However, the excellent results for the cm bands radio wave can be attributed to the extensive study from Goldstone Apple Valley Radio Telescope Observations \citep{Velusamy2020}. From their study,  the observations obtained made us believe that the radio emission originated in the chromosphere and corona; furthermore, they achieved to obtain the source brightness temperatures and angular sizes as a function of frequency. These results were made known in terms of gyroresonance mechanism across the active region of the sun.

Solar eclipse is a rare and attractive event occurring 2 or 5 times per year. Optical astronomers usually study the solar features happening at the photosphere, chromosphere and corona. We did combine radio observations for the 2019 and 2020 solar eclipses by making use 3 small radio telescopes in Hong Kong. The results were cross checked as to identify any discrepancy. The dates of observations were on 26th of December, 2019 and 21st of June, 2020, respectively. This paper will mainly discuss one of the most comprehensive observation result among those trials, others will be included in Appendix section.

  This study has three goals. First, to obtain the time variation of radio two-dimensional spatial mapping of solar eclipses. Second, to obtain total radio flux density time curves (light curve analogy in radio). Lastly, to verify if 21 cm radio band can reveal any physical properties of solar eclipses through analyzing historical worldwide radio eclipses. This paper is also a collection of historic radio observations in Hong Kong, which will make further studies of comparison between radio quiet zone in Hong Kong and that in other countries feasible. Italy, in which our telescopes' manufacturer Radio2space Company bases, would be the first on our list, as the same SPIDER 300A radio telescope system is also installed at their National Radio Observatory in Bologna.

\section{Observations}
\noindent\underline{Partial Solar Eclipse on June 21, 2020 (magnitude 0.89, max eclipsed area 86.08\%)}
\newline

  The radio observations of the eclipse were made on June 21, 2020 at the Ho Koon Astronomical Center (hereafter HKAC, Longitude: $114^o\;6'\;29.3076''E$, Latitude: $22^o\;23'\; 1.644'' N$, Altitude:149 m), Stanley Ho Astronomical Observatory (hereafter SHAO, Longitude: $114^o\; 13'\; 24.0414'' E$, Latitude: $22^o14'32.2362'' N$ , Altitude: 4.6 m), and Physics Building Dome of University of Hong Kong (hereafter HKU, Longitude: $114^o\;8'\; 23.262''E$, Latitude: $22^o\;13'\; 59.7'' N$, Altitude: 120 m), respectively.

  Small Radio Telescope (SRT), developed by MIT Haystack Observatory, was used in HKAC. SRT is centered at 1420 MHz with half-power beamwidth (HPBW) of 7 degrees. While SPIDER 300A radio telescope systems, developed by Radio2space Company, were used in both SHAO and HKU. The telescopes were optimized at 1420 MHz with HPBW of 4.03 degrees. The diameters of the telescopes that we used in SHAO, HKU and HKAC are 3m, 3m and 2.3m respectively with their corresponding bandwidths at 50MHz, 50MHz and 50kHz. The so-called ``on-the-fly two dimensional mapping" used at SHAO is indeed a mode of moving the antenna horizontally from left to right. Once the antenna completed mapping the first row, it moves downward and starts the succeeding row. This mode of movement of scanning horizontally and one row downward will continue and repeat until the designated area of sky had been fully scanned.

  And the ``on-off method" is that the antenna will be operated to scan the designated sky so that the feed horn is made to align to the target ``on mode" for a period of time, occasionally the feed horn is made to align to the ambient space ``off mode" for another period of time as to get the ambient signals. This ``on-off method" is adopted repeatedly during the observing time as to ensure to properly receive the signals as well as the ambient noises.

  Always on-source tracking method was used at HKAC and HKU, while on-the-fly two-dimensional mapping method was applied at SHAO, which allowed us to obtain the time variation mapping of radio sun during eclipse, which will then be presented as sequence diagrams in the later section. The sun was mapped with a $7\times7$ grid covering $10 \times 10$ degrees$^2$, and with step size of 1.6 degrees. Integration time for each grid was 1 second. All the data recordings were saved in the standard FITS format.

  Unfortunately, data taken at HKU is corrupted which made us impossible to compare with that from SHAO. Data of HKAC encountered some suspicious problems which is then put in appendix for further discussion. There was a desperate failure in tracking the sun. As a result, the data obtained from HKU was not entirely attributed from the target, so we considered the data corrupted. This paper will be mainly focused on SHAO data.

\section{Data reduction}
 Noted that the calibration system of SPIDER 300A is still under development at the moment of this manuscript being drafted, arbitrary unit and percentage changes were used during analysis.

   Since the telescope was scanning across the surface of the sun, the offsets between the telescope axis and the sun in azimuthal and altitude at each moment were continually varying. Those offset functions of time can be retrieved from fits headers. We can then scale the received power, which is diminishing along the off-axis distance, with respect to each channel according to the beam profile. The scaled powers were added together after flagging particular channels which were severely  polluted by the local interference. We assume the solar power spectrum smooth and without strong emissions in any particular frequency, so that flagging some channels will not significantly affect the result if we only consider the percentage changes in power throughout the eclipses.
  For our SPIDER 300A system is an affordable radio telescope system, the small deviations from the electronic components were inevitable. During the observation, we made use of different gain values for the both left and right hand polarization as to minimise the deviation.

  For the 2020 eclipse, the sun was observed by on-the-fly two-dimensional mapping method mentioned in the previous section. Since the angular distance between the sun and the measurement center varied from time to time throughout the scanning process, in order to obtain the true radio flux of the sun at each moment, the power diminishing effect along off-axis distance must be compensated according to the beam profile. The offset distances and beam parameters could be found in FITS, through dividing measured flux by beam respond at corresponding offset, time series of the power change will then be obtained. Power of the uneclipsed sun is estimated by averaging the data from $-6$ hr. to $-2$ hr. before the eclipse maximum.

  For comparison, the change of eclipsed area is also included in the plots noted as optical. Equatorial positions of the sun and the moon were calculated by Skyfield \citep{Rhodes2019}, and the area changed was calculated according to equations provided in Maplesoft webpage \citep{Jason2019}.

\section{Discussion}
For this study, we accomplished the 3 major goals mentioned in the Introduction section. Hence, for this discussion, we will make reference on these 3 goals to proceed with detailed consideration.
\newline

\noindent\underline{Time variation of radio two-dimensional spatial mapping of solar eclipses}
\newline

  The two-dimensional mapping false colour diagram was created by making use of the 1.42 GHz uncalibrated SPIDER 300A data in arbitrary units from 2020 eclipse at SHAO. The location for the SHAO is in the valley of the Tai Tam reservoir, where a hill blocks the west side of the sky, the last hour of 2020 eclipse was not recorded. Fortunately, the sequence diagrams demonstrating the first contact, partial phases and the maximum eclipse of the partial solar eclipse were obtained. Please be reminded that this was 0.89 partial solar eclipse instead of total solar eclipse.

\begin{figure}[h!]
\begin{center}
\includegraphics[width=.7\textwidth,angle=-0]{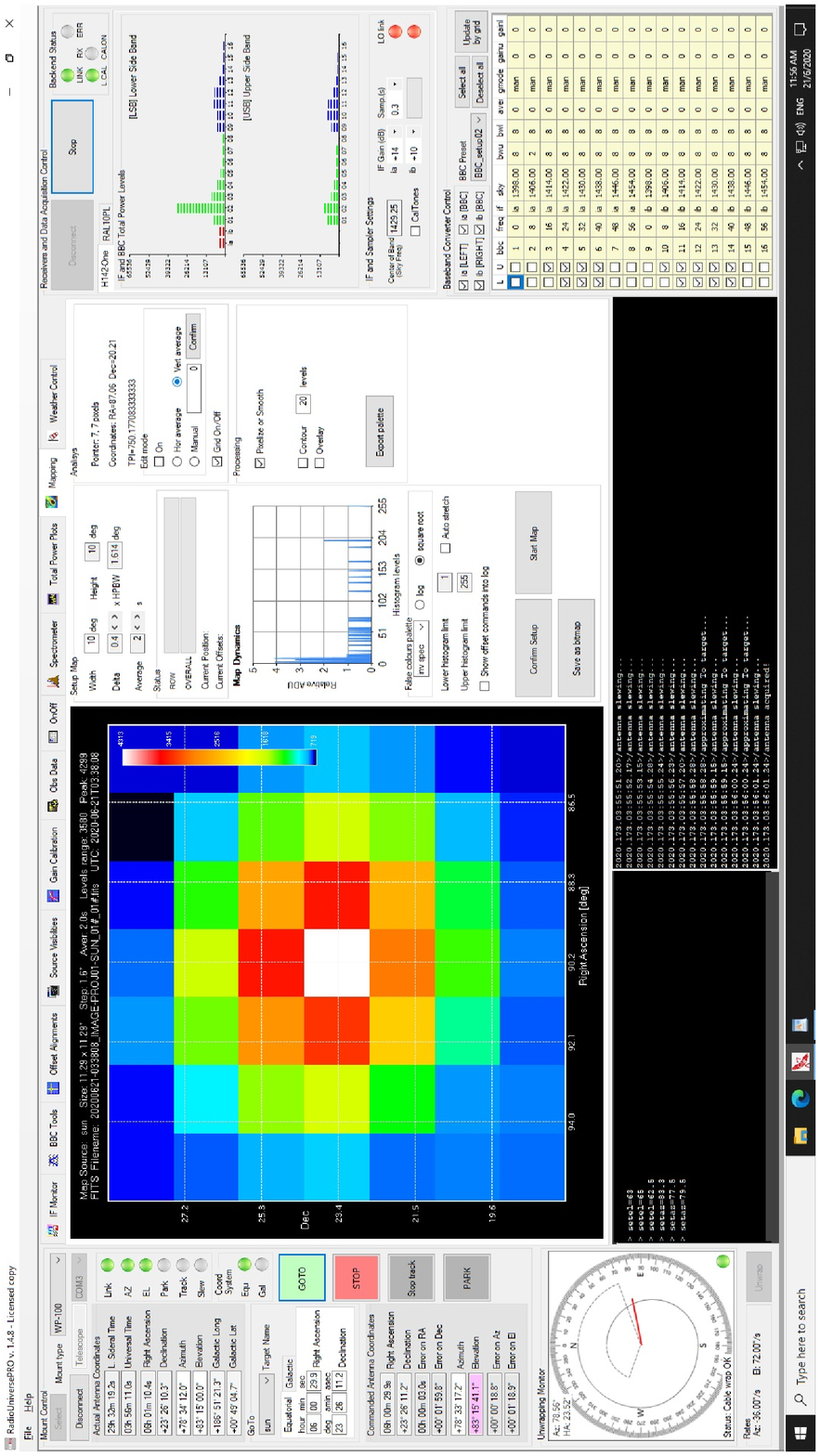}
\newline 
\caption{The standard interface for the RadioUniversePRO software shows the sun in raw data format.} \label{fig01}
\end{center}
\end{figure}

\begin{figure}[h!]
\begin{center}
\includegraphics[width=0.5\textwidth,angle=0]{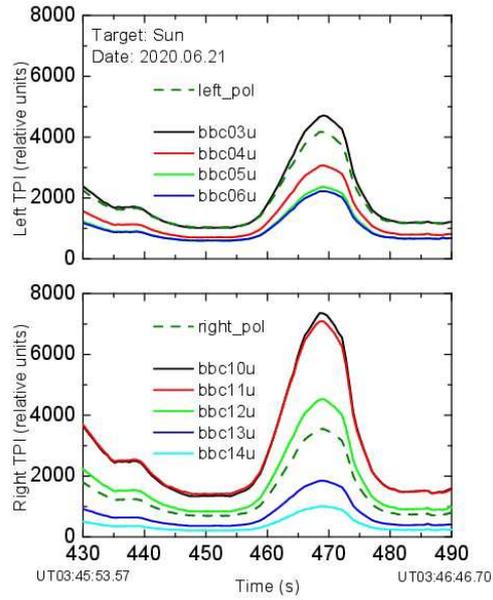}
\caption{The real time flux density curve versus time plots. The plots represent dual polarization for the sun's signal} \label{fig03}
\end{center}
\end{figure}

  The mapping result is shown in Fig. \ref{fig01}. Fig. \ref{fig01} shows a mapping diagram before the solar eclipse. The UT that we made the figure was at 03:56. We adopted 2s for the integration time and a step size is of 1.614 degree for each pixel. Fig. \ref{fig01} was obtained before the solar eclipse. It was included as to give a proper comparison to the later figure obtained during the eclipse.  Fig. \ref{fig03} shows the real time dual circular polarization time plots generated from the built-in software RadioUniversePRO v.1.4.8. Fig. \ref{fig03} shows schematically a part of real time intensity changes during mapping as the telescope beam scanning across the sun. The intensity was found to vary as shown in Fig. \ref{fig03}.

\begin{figure}[h!]
\begin{center}
\includegraphics[width=0.5\textwidth,angle=-90]{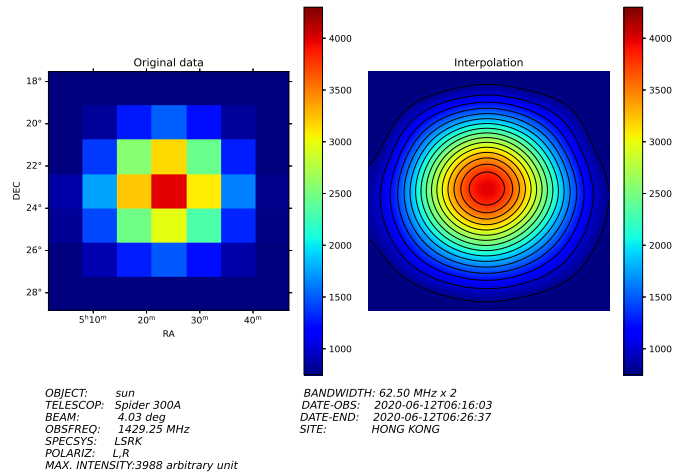}
\caption{The usual daily mapping for the sun by making use SPIDER 300A} \label{fig04}
\end{center}
\end{figure}

\begin{figure}[h!]
\begin{center}
\begin{tabular}{cc }
\includegraphics[width=0.4\textwidth,angle=-90]{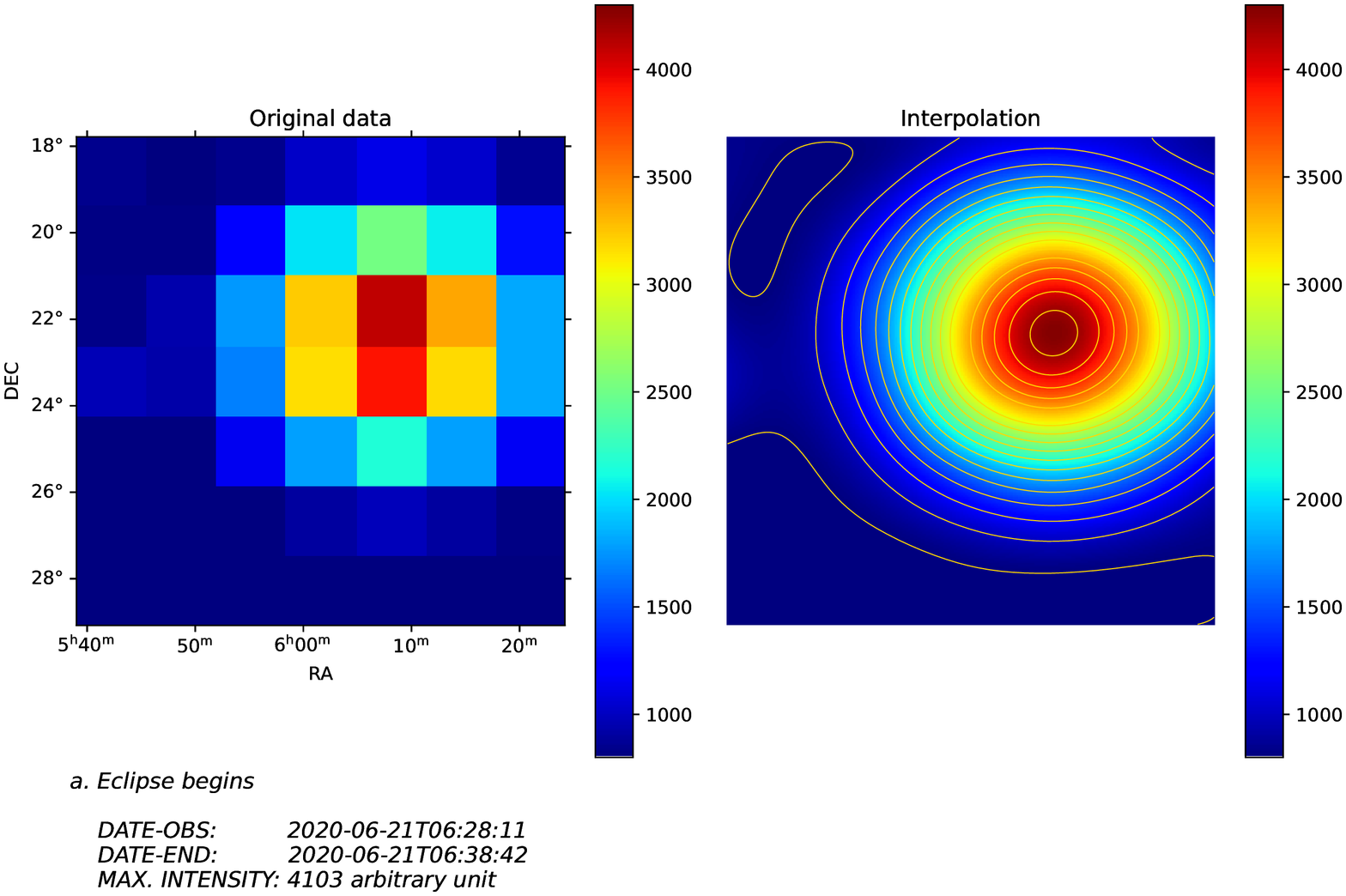}&
\includegraphics[width=0.4\textwidth,angle=-90]{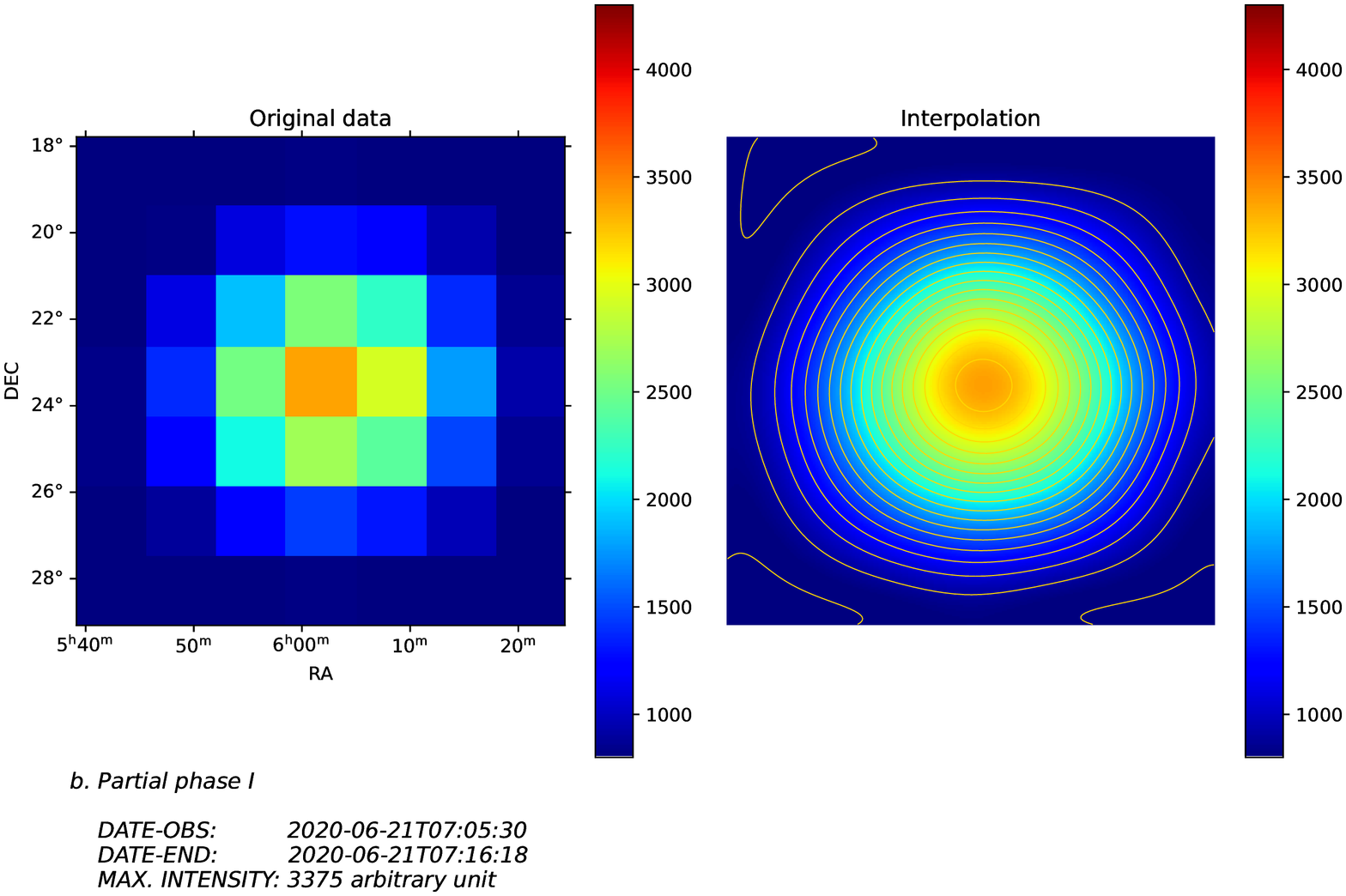}\\
\includegraphics[width=0.4\textwidth,angle=-90]{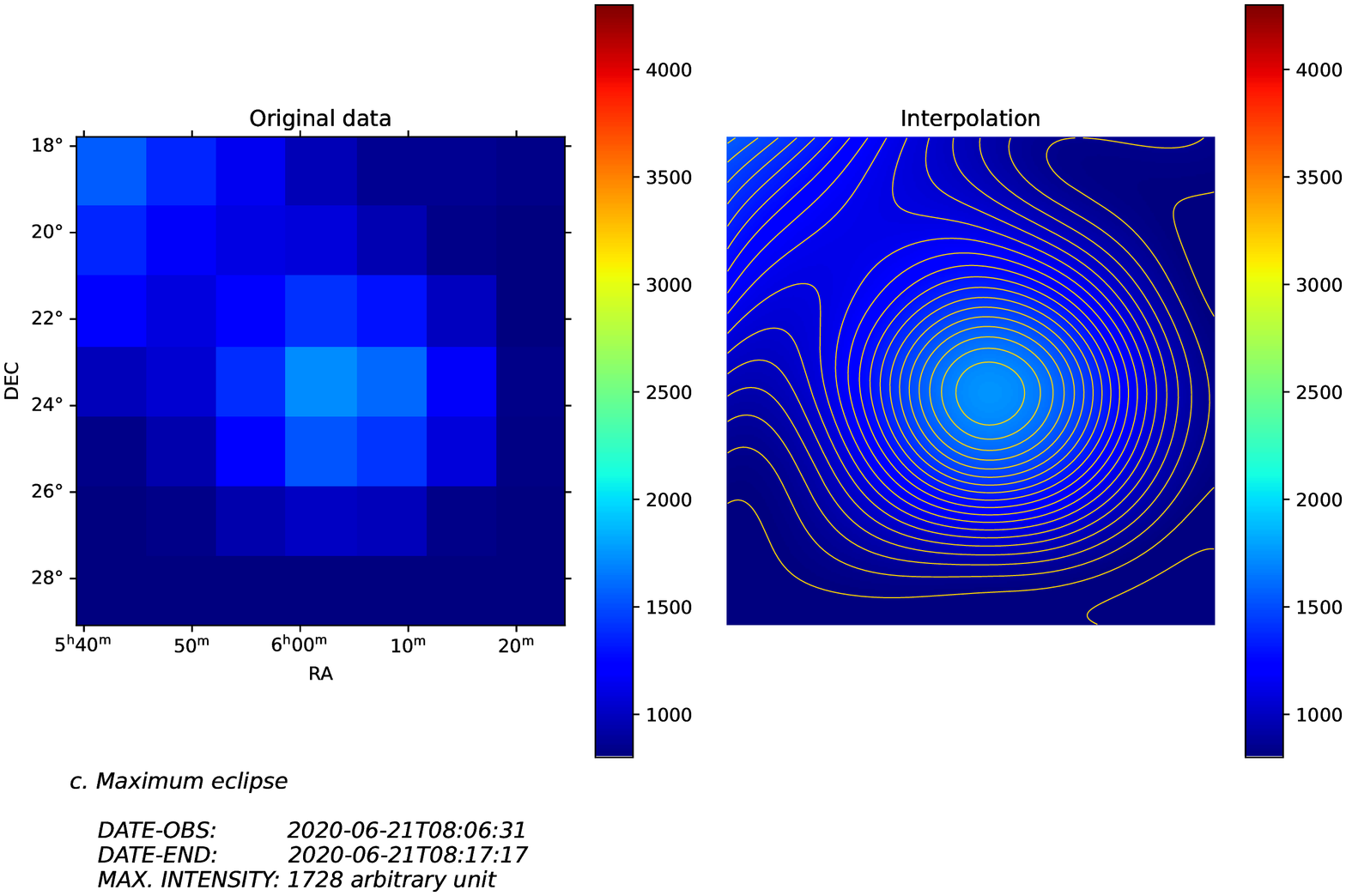}&
\includegraphics[width=0.4\textwidth,angle=-90]{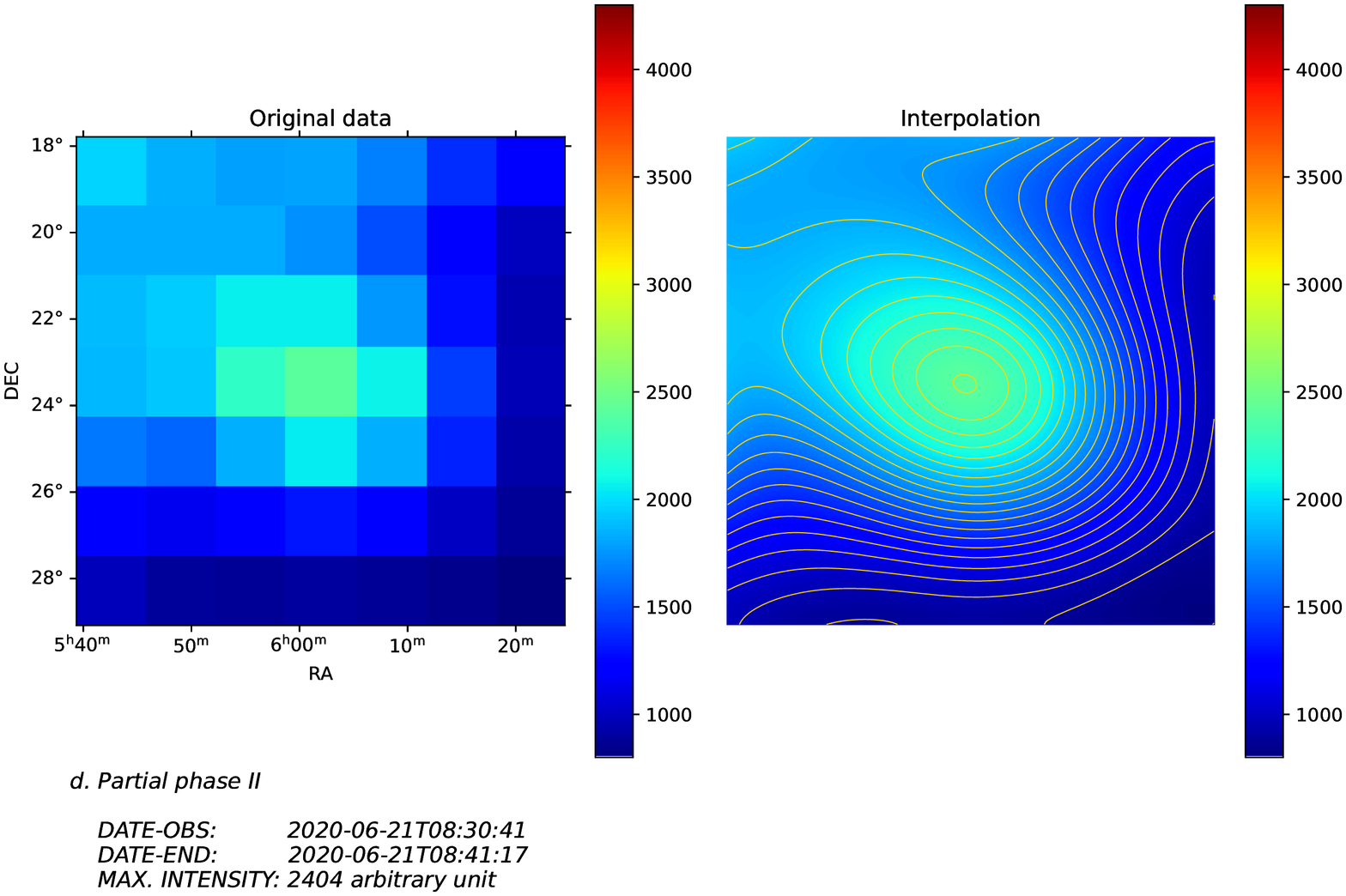}
\end{tabular}
\includegraphics[width=0.4\textwidth,angle=-90]{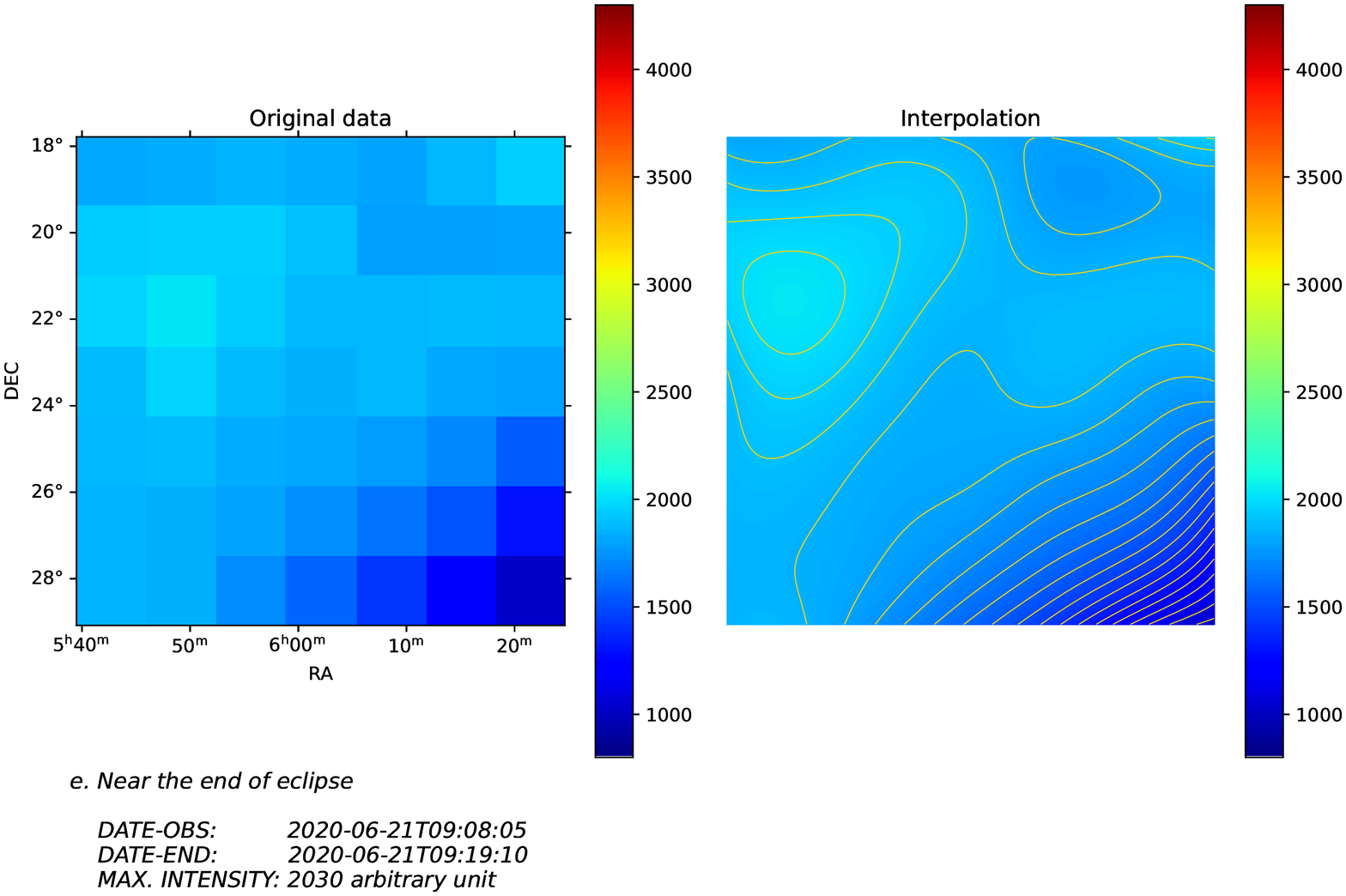}
\caption{Sequence diagrams of the 0.89 partial solar eclipse in Hong Kong on June 21, 2021: a. Eclipse begins, b. Partial phase I, c. Maximum eclipse, d. Partial phase II, and e. Near the end of eclipse (blocked by hill) } \label{fig05}

\end{center}
\end{figure}

   We would admit that it took roughly 15 minutes for a scanning and the follow-up scanning which included the mechanical movement by the antenna, setting of the device parameters, the fine tuning during the observation, recording the data, and the resetting for the follow-up scanning. New mapping with the same setting was repeated immediately after the previous scan. The whole process continued until the sun was blocked by the hills at west of SHAO. A sequence of $10 \times10$ degrees$^2$ images of the eclipsed sun was then obtained. Fig. \ref{fig04} shows an example for the uneclipsed sun that observed in some other day for comparison. Fig. \ref{fig05} shows  a sequence of diagrams as to present the different stages of the eclipse.

   Fig. \ref{fig05} clearly shows that the received power of the sun decreases as the eclipsed area increases. Since the beam of the telescope is 4.03 degrees, which is much larger than that of the sun ($<$1 degree); unlike what we expect in optical images, the radio morphology of the sun remains circular throughout the process.

  The latter half of the sequence diagrams of the eclipse demonstrates the effect from the blockage of the hills nearby. Although our sequence diagrams are not able to cover the whole eclipse from the start to the end, the moment of the maximum eclipse (frames from UT06:28 to UT09:19 in Fig. \ref{fig05} ) is recorded, enabling further analysis of the process from the beginning to the maximum of radio eclipse. For interpolation of Fig. \ref{fig05}, we use ``RectBivariateSpline" function from SciPy package to interpolate data over rectangular meshes by bivariate spline approximation. In our actual operation, 36 values are interpolated between meshes in each direction, with 3 degrees of the bivariate spline. More comprehensive radio solar eclipse observations from better sites with better occasions and better instrumental conditions in coming years are expected, our result will serve as one of those records in Asia for future studies.
\newline

\noindent\underline{Total radio flux density time curves}
\newline

\begin{figure}[h!]
\begin{center}
\includegraphics[width=0.7\textwidth]{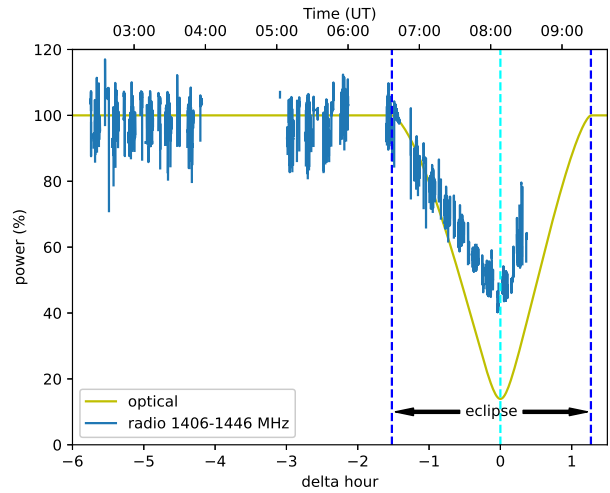}
\caption{The percentage change of radio power during the partial solar eclipse on June 21, 2020 (magnitude 0.89, max eclipsed area 86.08\%). The Y-axis represents the flux density (power) percentage change and X-axis represents the corresponding time in both delta hour and UT scale.)} \label{fig06}
\end{center}
\end{figure}

  Based on the data processing described, the percentage change of received radio power during the eclipse is shown in Fig. \ref{fig06} for 2020 partial eclipse.

  As the resolution of the radio observation is less than that  of the optical, previous studies \citep{Sherwood1978, Tan2009} showed that the radio sun in 21 cm appeared larger than that in optical, wider and shallower dips of radio power change compared to that of optical are to be expected. Our results from 2020 eclipse agree with the expected demonstrating shallower dip when compare with the optical eclipse models. For SHAO 2020 data, although we covered the very beginning of the eclipse, no significant delay is found from the reduced data. We postulate the delay may be too small due to the almost perpendicular intersecting angle between the sun and the moon such that our instrument was not sensitive enough to determine the beginning moment of the radio eclipse. The expected delay mentioned is reasonably attributed to different paths, and so different thickness of the atmosphere of the optical and radio waves taken passing through between the almost perpendicular angle and later more ``inclined angle" for the Sun's positions appeared in the sky. And no significant delay was found. If more data at around $-2$ hour before eclipse maximum were obtained, a better estimation might be achieved.

\begin{figure}[h!]
\begin{center}
\includegraphics[width=0.7\textwidth]{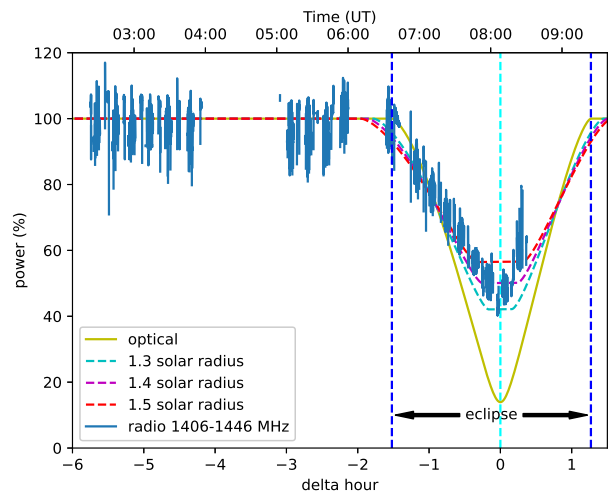}
\caption{ The simulated light curve for the 2020 partial solar eclipse with different radio sun's radius.} \label{fig07}
\end{center}
\end{figure}

  From the light curve, we found the drop in radio power is 55$\pm$ 5\% at maximum eclipse, which means that the radio radius of the Sun is larger than that in optical. We can estimate the ratio between radio solar radius and that of optical from their differences in power drop. the simulation result is shown in Fig. \ref{fig07}.  Assuming the radio behavior of the sun is similar to that in optical, the morphology of radio pattern is basically symmetric circular disc. We can obtain the expected light curves by substituting different solar radii(ranging from 1 to 1.5 solar radius) in the equation \citep{Jason2019}. The most fitted model would indicate the measured solar radius of our observations.

During the solar eclipse, the eclipsed area in optical was smaller  than that of the radio signals. Therefore, the power drop of the signals in optical would be more than that of the radio.  And the result was shown in Fig. \ref{fig07}.

  Solar radius is scaled with specific factors to see how the curves change. From the above result, we estimated that the radio solar radius is roughly 1.4$\pm$0.1 of that in optical. \citep{Sherwood1978, Tan2009}

\section{Conclusion}
Based on the above observational results, we were glad to have accomplished the original planned goals for this study. Firstly, we managed to obtain the two-dimensional radio mapping animation for the 0.89 partial solar eclipse on 21 June 2020 in Hong Kong.  The radio eclipsed area in 2020 partial solar eclipse as observed was different and not in proportion to that of the optical as expected. There was about 55 $\pm$ 5\% of the maximum radio eclipse recorded compared to a 86.08\% eclipsed area in optical. And the radio solar radius detected during the solar eclipse in 21 cm was related to chromosphere and corona, but not to the photosphere. The photosphere of the sun defines the solar disc optically. During the solar eclipse, the optical light of the sun is blocked by the moon. The radio signals in 21 cm at the chromosphere and the corona of the sun, however, remain observable during the solar eclipse. Therefore, we may recognise the signals observed as the radio solar disc.

The 2020 observation for the 21 cm radio solar eclipse was the unprecedented in Hong Kong. Probably, other South East Asia countries may have similar observations. Therefore, our observation data obtained can serve effectively to contribute to the radio data all over the world.

There was a radio quiet zone in Hong Kong, such that we propose constructing a  radio interferometer array over there. In fact, we wish to implement the plan soon since it is reasonably good for the development in radio astronomy in Hong Kong.

And we developed a simple systematic pipeline data reduction approach for dealing  with the data obtained from the SPIDER 300A telescope. And the calibration function for the SPIDER telescope is expected to be completed soon, and hence we can properly calibrate all the coming observational results.

The results we have obtained are good enough for initiating further studies and collaborations with other countries and research groups, and for comparing our data with their results.

\begin{acknowledgements}
We would like to express our deep gratitude to the late Dr. Stanley Ho and his family for their generous donation and huge support to the Stanley Ho Astronomical Observatory at Tai Tam, Hong Kong, who provided the radio-telescope and all accessories required for this solar eclipse observation and data acquisition. Special thanks are given to Prof. Yuen Kwok Yung as facilitator and Daisy Ho as donation coordinator and donor.
\end{acknowledgements}

\makeatletter
\def\@biblabel#1{}
\makeatother

\newpage
\section{Figures}

\begin{figure}[h!]
\begin{center}
\includegraphics[width=0.7\textwidth]{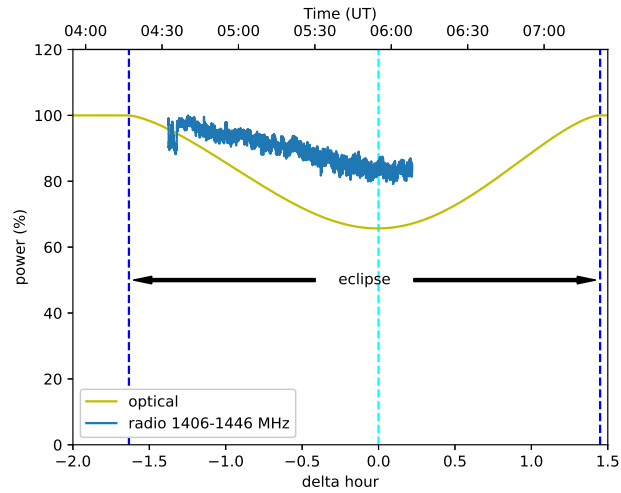}
\caption{ The percentage change of radio power during the partial solar eclipse on December 26, 2019 (magnitude 0.46, max eclipsed area 34.32\%)} \label{fig08}
\end{center}
\end{figure}

\begin{figure}[h!]
\begin{center}
\includegraphics[width=0.7\textwidth]{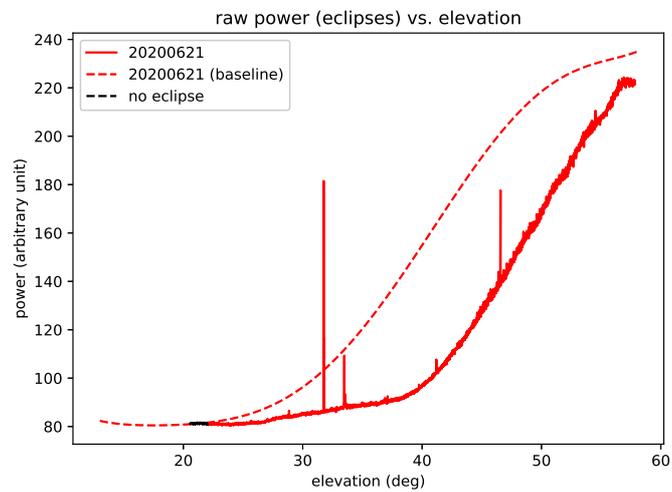}
\caption{ Raw data in power vs. elevation measured by SRT in HKAC. Solid line represents the eclipsed data, dotted line is the baseline for calibration.} \label{fig09}
\end{center}
\end{figure}

\begin{figure}[h!]
\begin{center}
\includegraphics[width=0.7\textwidth]{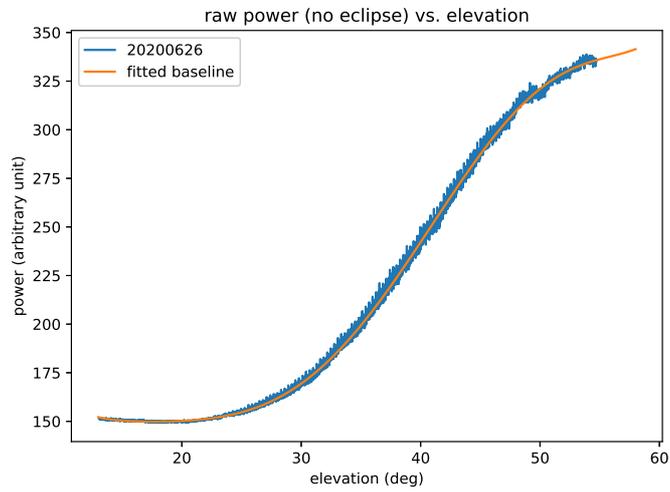}
\caption{Raw data of uneclipsed sun measured by SRT in HKAC} \label{fig10}
\end{center}
\end{figure}

\begin{figure}[h!]
\begin{center}
\includegraphics[width=0.7\textwidth]{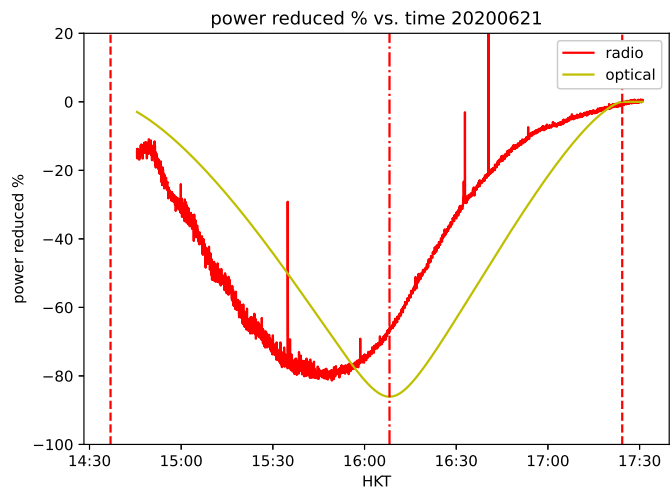}
\caption{ Reduced radio eclipse data by SRT in HKAC shown according to Hong Kong local time (HKT): UTC+8h} \label{fig11}
\end{center}
\end{figure}

\begin{figure}[h!]
\begin{center}
\includegraphics[width=0.7\textwidth,,angle=-90]{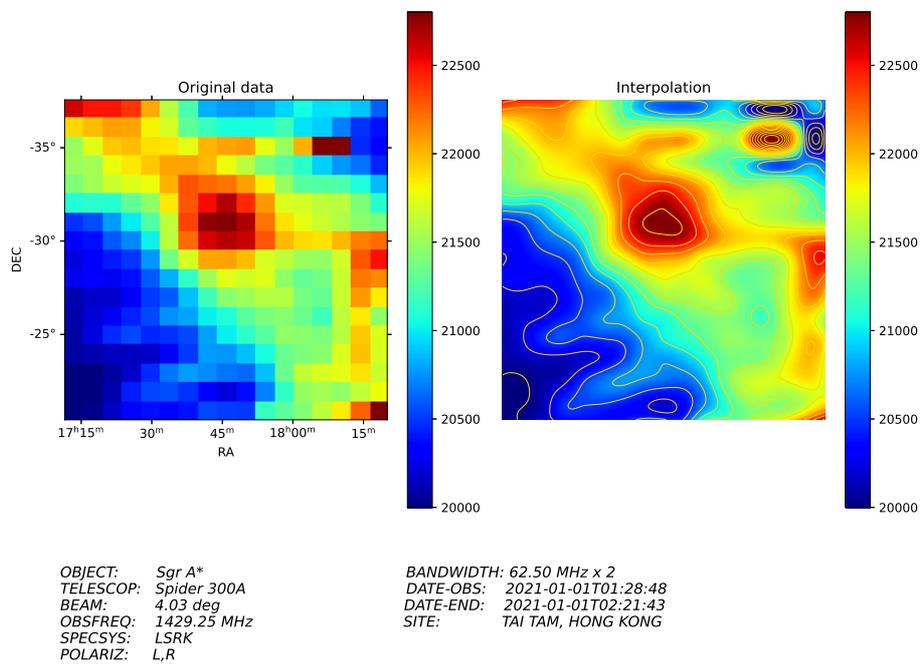}
\caption{ Radio mapping of the Sagittarius A* after processing} \label{fig13}
\end{center}
\end{figure}

\newpage

\appendix
\section{}
  The same set of SPIDER 300A telescope is also used in 2019 partial solar eclipse observation in Hong Kong. The details will be discussed here.

  The radio observations of the eclipse were made on December 26 at the HKAC, SHAO, and HKU, respectively. The magnitude of the eclipse is 0.46, indicating that the maximum eclipsed area is 34.32\%. SPIDER 300A radio telescope systems were used in both SHAO and HKU. Always on-source tracking method was used for all observations. Unfortunately, data collected at HKU is corrupted, only the result from SHAO will be discussed in this section, namely SHAO 2019.

  Different from that of 2020 eclipse, since the telescope was always tracking at the sun, we can simply add up all data from different channels covering the interested frequency range to obtain time series of the flux change. As the instrument was not calibrated, measurement of uneclipsed sun is preferred for estimating the percentage of power change during eclipse. Unfortunately, this observation only covered the first half of the event, we have no clue about the total flux of the uneclipsed sun measured by the same set of instruments. At the best we could do, we take the maximum power of the time series as an educated or reasonable guess of the power of the uneclipsed sun, which inevitably did introduce systematic error but at a possible minimum level. The result is presented in Fig. \ref{fig08}.

   From the analysis, the SHAO 2019 data show a shallower dip (18$\pm$ 5\%) when compare with that in optical (34.32\%). Since it did not cover the very beginning of the eclipse, we cannot tell whether radio eclipse begin earlier than optical eclipse from this data set.
\section{}
 As mentioned in observation section, HKAC also participated in the 2020 eclipse observation. While SPIDER 300A record data by  on-the-fly method in SHAO, SRT, used in HKAC, only record the raw readings from the detector, of which the sky background must be accounted before data analysis ( so called ``on-off mode" ). The raw eclipsed data are presented in Fig. \ref{fig09} as the red solid line.

   We cannot tell when the sun is eclipsed without further data reduction. In order to obtain the radio power difference between eclipsed and uneclipsed sun measured by the SRT at HKAC, another measurement of the sun was made 5 days later, as shown in Fig. \ref{fig10}. It provided us the variation of solar radio power along solar elevation measured by our SRT. The data is then approximated by fitting a polynomial as shown as orange solid line in Fig. \ref{fig10} and dotted line in Fig. \ref{fig09}. By comparing the difference between two curves, atmospheric factors are eliminated, and variation of the eclipse can be revealed as shown in Fig. \ref{fig11}.

   The reduced data shows a reasonable dip indicating a typical eclipse event. However, our data also show a large time shift between radio data and optical model. We have no clue of the reasons behind the shift. It could be instrumental errors or methodological errors. We tried everything to identify the sources of error, but it did not help. We note this data as highly suspicious and hope more insight would be offered in the future.

\section{}

  We collected many interesting signals in Hong Kong sky which contributed to the historical detection in 21cm line in Hong Kong. Fig. \ref{fig13} show the radio mapping of the galactic center, also known as Sagittarius A$^*$.

\received{\it *}
\end{document}